\documentclass[twocolumn,notitlepage,nofootinbib]{revtex4-2}
\usepackage{amsmath,amssymb,bm,graphicx,hyperref}
\allowdisplaybreaks[1]

\renewcommand\d{\partial}
\newcommand\grad{\bm{\nabla}}
\newcommand\0{\bm{0}}
\newcommand\A{\bm{A}}
\newcommand\B{\bm{B}}
\newcommand\E{\bm{E}}
\renewcommand\j{\bm{j}}
\renewcommand\k{{\bm{k}}}
\renewcommand\r{\bm{r}}
\newcommand\x{\bm{x}}
\newcommand\y{\bm{y}}
\newcommand\z{\bm{z}}
\newcommand\C{\mathcal{C}}
\renewcommand\P{\mathcal{P}}
\newcommand\m{\mathrm{m}}
\let\Re\relax\DeclareMathOperator\Re{Re}

\begin{document}

\title{Chiral light amplifier with pumped Weyl semimetals}

\author{Yusuke Nishida}
\affiliation{Department of Physics, Tokyo Institute of Technology,
Ookayama, Meguro, Tokyo 152-8551, Japan}

\date{December 2022}

\begin{abstract}
Parallel electric and magnetic fields applied to Weyl semimetals pump axial charge via the axial anomaly until balanced by intervalley relaxation.
The resulting nonequilibrium steady state exhibits the chiral magnetic effect as well as the anomalous Hall effect, which coupled with Maxwell's equations leads to unstable electromagnetic waves at low frequency and long wavelength.
Here, we show that such chiral magnetic instability manifests itself as anomalous reflectance of the surface of pumped Weyl semimetal.
Depending on electric, chiral magnetic, and anomalous Hall conductivities, the reflectance is found to exceed unity in a finite range of frequency for a circularly polarized light incident along the direction of Weyl node separation.
\end{abstract}

\maketitle
%\tableofcontents

\section{Introduction}
\begin{figure}[b]
\includegraphics[width=0.85\columnwidth]{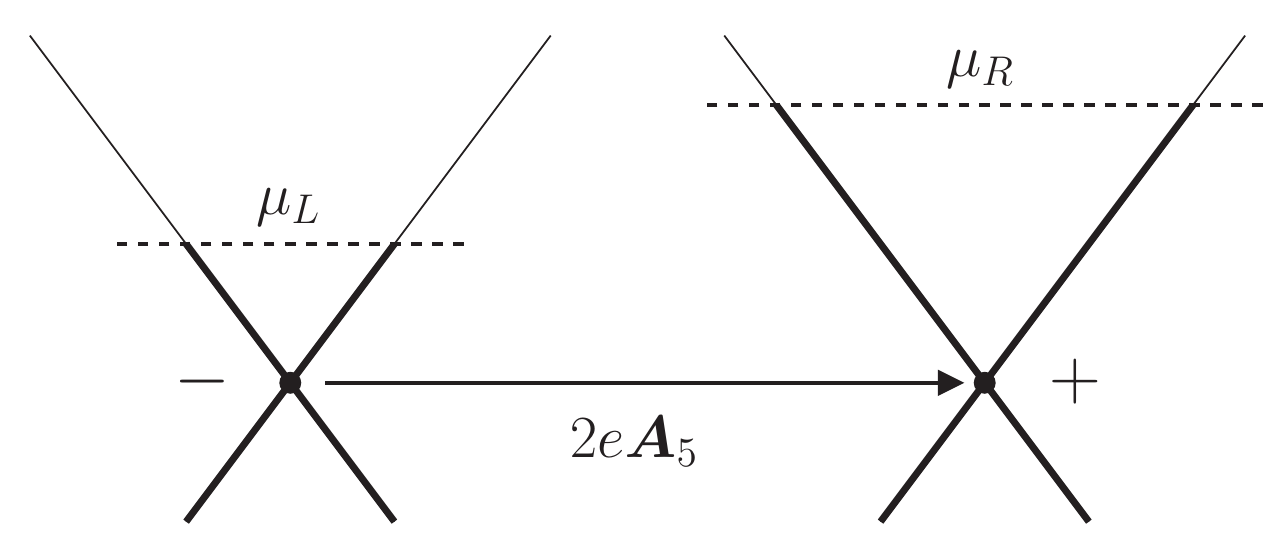}
\caption{\label{fig:nodes}
Two Weyl nodes with chirality $\chi=\pm$ separated in momentum direction by $2e\A_5$.
The axial chemical potential $\mu_5=(\mu_R-\mu_L)/2$ can be nonzero in nonequilibrium steady states.}
\end{figure}

Weyl semimetals constitute a new class of topological materials in three dimensions, where the valence and conduction bands touch each other at isolated points, called Weyl nodes, in the Brillouin zone~\cite{Murakami:2007,Wan:2011}.
Among others, they offer exciting platforms hosting a variety of anomalous transport phenomena~\cite{Armitage:2018,Gorbar-Miransky-Shovkovy-Sukhachov}.
One such example is the anomalous Hall effect,
\begin{align}\label{eq:AHE}
\j_\mathrm{AHE} = -\frac{e^3}{2\pi^2\hbar^2}\A_5\times\E,
\end{align}
predicting the electric current perpendicular to an applied electric field without a magnetic field~\cite{Yang:2011,Burkov:2011}.
The anomalous Hall effect is possible when the time-reversal symmetry is broken so that a pair of Weyl nodes with opposite chiralities is separated in momentum space by $2e\A_5$ as depicted in Fig.~\ref{fig:nodes}.

Another example is provided by the chiral magnetic effect,
\begin{align}\label{eq:CME}
\j_\mathrm{CME} = \frac{e^2}{2\pi^2\hbar^2}\mu_5\B,
\end{align}
predicting the electric current parallel to an applied magnetic field in the presence of an axial chemical potential $\mu_5$~\cite{Kharzeev:2008,Fukushima:2008}.
Because $\mu_5$ vanishes in equilibrium, the chiral magnetic effect can be activated only by driving the system out of equilibrium~\cite{Vazifeh:2013,Basar:2014,Landsteiner:2016}.
In particular, by applying parallel electric and magnetic fields~\cite{Son:2013,Burkov:2014}, possible signatures of the chiral magnetic effect have been experimentally observed in Weyl semimetals~\cite{Armitage:2018,Gorbar-Miransky-Shovkovy-Sukhachov}.
The nonequilibrium nature of the chiral magnetic effect can also be understood by coupling Eq.~(\ref{eq:CME}) with Maxwell's equations~\cite{Shovkovy:2022}.
The resulting collective excitations at low frequency and long wavelength turn out to be unstable and grow exponentially over time as a consequence of positive feedback of the chiral magnetic effect to Amp\`ere's circuital law~\cite{Akamatsu:2014}.
Such chiral magnetic (or plasma) instability has been studied extensively with diverse implications for the early Universe and the quark-gluon plasma~\cite{Joyce:1997,Giovannini:1998,Boyarsky:2012,Tashiro:2012,Akamatsu:2013,Akamatsu:2014,Tuchin:2015,Manuel:2015,Hirono:2015,Gorbar:2016}.

The question we ask in this Letter is how the chiral magnetic instability manifests itself now in Weyl semimetals.
Our answer proposed below as a matter of principle is anomalous reflection amplifying a circularly polarized light by the surface of Weyl semimetal with nonzero axial chemical potential.
We note that the electron charge is denoted by $e=-|e|$ throughout this Letter.

\section{Chiral magnetic instability}
Low-energy effective description of Weyl semimetals is provided by a pair of right- and left-handed Weyl fermions.
Their charge imbalance $\rho_5(t,\r)=\rho_R(t,\r)-\rho_L(t,\r)$ is called the axial charge and obeys the anomalous continuity equation of
\begin{align}
\frac{\d\rho_5(t,\r)}{\d t} + \grad\cdot\j_5(t,\r)
= \frac{e^3}{2\pi^2\hbar^2}\E(t,\r)\cdot\B(t,\r) - \frac{\rho_5(t,\r)}{\tau_5}.
\end{align}
Here, the first term on the right-hand side is the celebrated axial anomaly violating the local conservation of axial charge in the presence of electric and magnetic fields~\cite{Adler:1969,Bell:1969,Nielsen:1983}, whereas the second term is added to phenomenologically incorporate the relaxation time $\tau_5$ due to intervalley scatterings~\cite{Son:2013}.
When uniform and static fields are applied, the axial charge is pumped until the axial anomaly is balanced by the intervalley relaxation so that the nonequilibrium steady state is reached with
\begin{align}
\rho_5 = \frac{e^3\tau_5}{2\pi^2\hbar^2}\E\cdot\B.
\end{align}
The nonzero axial charge then leads to the nonzero axial chemical potential $\mu_5=(\mu_R-\mu_L)/2$ as depicted in Fig.~\ref{fig:nodes}, provided that fermions equilibrate in each valley due to intravalley scatterings with the relaxation time of $\tau\ll\tau_5$.
Although the axial chemical potential may be spacetime dependent, it is hereafter considered to be a fixed constant and nonequilibrium steady states with $\mu_5\neq0$ are referred to as ``pumped Weyl semimetals.''
We note that the axial charge can also be pumped transiently by distorting a crystal with strain~\cite{Cortijo:2016} or by heating Weyl nodes unequally under a strong electric field~\cite{Nandy:2020}.

Such a pumped Weyl semimetal subjected to electric and magnetic fields develops an electric current of
\begin{align}\label{eq:current}
\j(t,\r) = \sigma_E\E(t,\r) + \sigma_M\B(t,\r) - \sigma_H\hat\A_5\times\E(t,\r).
\end{align}
Here, the first term on the right-hand side is the Ohmic current, the second term is the chiral magnetic current in Eq.~(\ref{eq:CME}), and the third term is the anomalous Hall current in Eq.~(\ref{eq:AHE}), which can be derived in the framework of consistent chiral kinetic theory with the relaxation-time approximation~\cite{Gorbar:2017,Amitani:2023}.
Their conductivities with $v_F$ being the Fermi velocity are provided by $\sigma_E=[e^2\tau/(6\pi^2\hbar^3v_F)][2\pi^2(k_BT)^2/3+\mu_R^2+\mu_L^2]$, $\sigma_M=e^2\mu_5/(2\pi^2\hbar^2)$, and $\sigma_H=e^3|\A_5|/(2\pi^2\hbar^2)$, respectively, whose spacetime nonlocality can be neglected for sufficiently slow external fields with their frequency and wave vector satisfying $|\omega|,v_F|\k|\ll1/\tau\ll k_BT/\hbar$, $\mu_R/\hbar$, $\mu_L/\hbar$~\cite{Amitani:2023,Amitani:footnote}.

The dynamics of electromagnetic fields can be studied by coupling Eq.~(\ref{eq:current}) with Maxwell's equation,
\begin{align}
\grad\times\B(t,\r) - \frac1{v^2}\frac{\d\E(t,\r)}{\d t} = \mu\,\j(t,\r),
\end{align}
where $v$ is the speed of light in a medium and $\mu$ is its permeability (not to be confused with the chemical potential).
Possible dispersion relations of electromagnetic waves can be found by solving the characteristic equation for a plane wave $\A(t,\r)=\A_{\omega,\k}\,e^{-i\omega t+i\k\cdot\r}$ in the temporal gauge for the sake of convenience.
Its analytic solutions were obtained at the linear order in $\omega\sim|\k|\ll\sigma$~\cite{Amitani:2023}, where one of them turns out to have a positive imaginary part in frequency for any wave vector.
Such unstable modes are nonpropagating if $\k\perp\A_5$ or $\A_5=\0$ with null Poynting vector, but they otherwise become propagating waves oriented to the direction of $\mu_5\A_5$, indicating anisotropic generation of electromagnetic waves from pumped Weyl semimetals.
An important insight from Ref.~\cite{Amitani:2023} is that, whereas the instability is caused by the chiral magnetic effect, it is the anomalous Hall effect that causes the propagation.
Therefore, we focus on the representative case of $\k\parallel\A_5$, where the characteristic equation simply reads
\begin{align}
k^2 - \frac{\omega^2}{v^2}
= i\bar\sigma_E\omega + \bar\sigma_Mk - \bar\sigma_H\omega
\end{align}
for circularly polarized
\begin{align}\label{eq:potential}
\A(t,\r) = A_{\omega,k}^\lambda(\hat\x + \lambda i\hat\y)\,e^{-i\omega t+ikz}
\quad (\lambda=\pm).
\end{align}
Here, $\lambda$ is the helicity of light and we employ $\E(t,\r)=i\omega\A(t,\r)$ and $\B(t,\r)=\lambda k\A(t,\r)$ as well as $\bar\sigma_E\equiv\mu\sigma_E$, $\bar\sigma_M\equiv\lambda\mu\sigma_M$, and $\bar\sigma_H\equiv\pm\lambda\mu\sigma_H$ depending on the direction of $\hat\A_5=\pm\hat\z$.

The chiral magnetic instability of a propagating electromagnetic wave can be understood from a complementary perspective by expressing its wave number in terms of frequency $\omega>0$ as
\begin{align}\label{eq:dispersion}
k &= \frac{\bar\sigma_M}{2} \pm \sqrt{\left(\frac{\bar\sigma_M}{2}\right)^2
+ i\bar\sigma_E\omega - \bar\sigma_H\omega + \frac{\omega^2}{v^2}} \\[2pt]
&= \label{eq:expansion}
\begin{cases}
\,\dfrac{\bar\sigma_M^2 - \bar\sigma_H\omega + i\bar\sigma_E\omega}
{\bar\sigma_M} + O(\omega^2), \smallskip\\
\,\dfrac{\bar\sigma_H - i\bar\sigma_E}{\bar\sigma_M}\omega + O(\omega^2).
\end{cases}
\end{align}
The low-frequency limit is presented by Eq.~(\ref{eq:expansion}) for the sake of clarifying underlying physics, where the upper sign in Eq.~(\ref{eq:dispersion}) corresponds to the upper (lower) solution in Eq.~(\ref{eq:expansion}) for $\bar\sigma_M>0$ ($<0$).
The upper solution in Eq.~(\ref{eq:expansion}) has the same sign in its real and imaginary parts so that it describes a usual electromagnetic wave in conducting media damping exponentially toward the direction of its propagation.
However, the lower solution in Eq.~(\ref{eq:expansion}) has the opposite signs in its real and imaginary parts for $\bar\sigma_H>0$ so that it describes an anomalous electromagnetic wave growing exponentially toward the direction of its propagation.
Such an unstable mode exists in a finite range of $0<\omega<\omega_0$ as indicated by Eq.~(\ref{eq:dispersion}), for which both chiral magnetic and anomalous Hall conductivities are essential.
Here, $\omega_0$ is defined by positive $\omega$ solving $\Re[\sqrt{(\bar\sigma_M/2)^2+i\bar\sigma_E\omega-\bar\sigma_H\omega+\omega^2/v^2}]=|\bar\sigma_M/2|$ for $\bar\sigma_H>0$.

\begin{figure}[t]
\includegraphics[width=0.8\columnwidth]{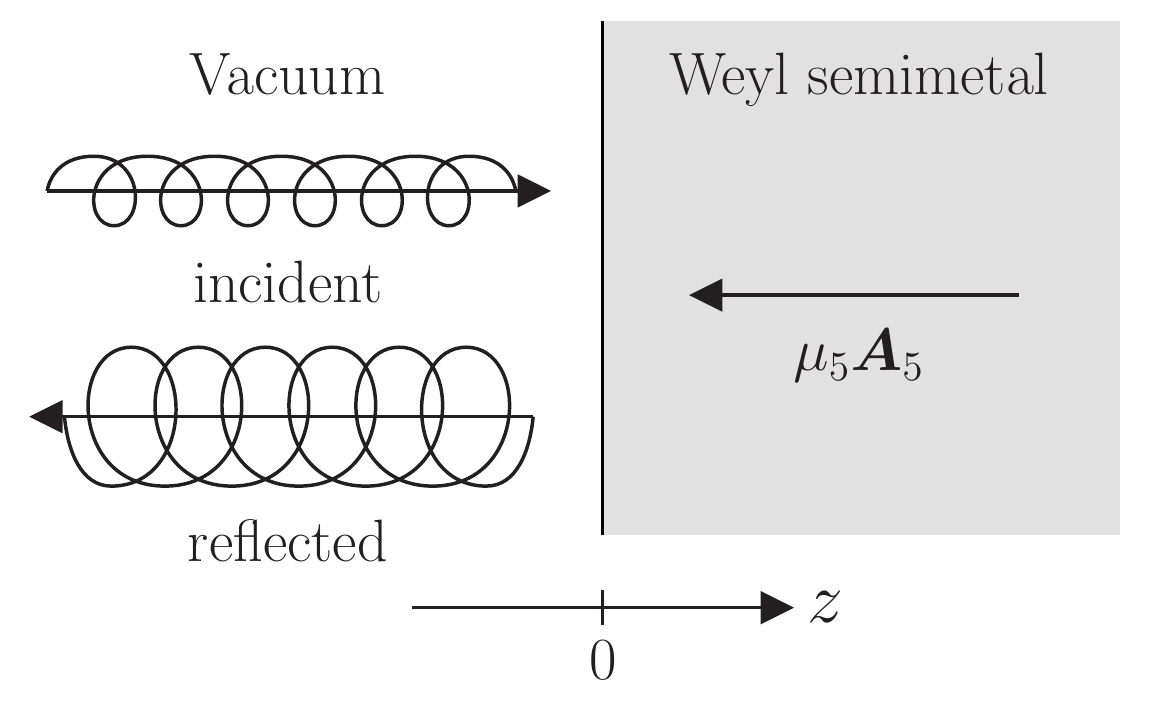}
\caption{\label{fig:setup}
Proposed setup where a circularly polarized light incident along the direction of Weyl node separation $\A_5$ is reflected by the surface of pumped Weyl semimetal with $\mu_5\neq0$.
The incident wave with selected helicity is to trigger the chiral magnetic instability generating an unstable electromagnetic wave propagating toward $\mu_5\A_5$.}
\end{figure}

\section{Anomalous reflectance}
We now propose that the chiral magnetic instability manifests itself as anomalous reflectance of the surface of pumped Weyl semimetal with $\mu_5\neq0$.
Our system consists of vacuum at $z<0$ and the Weyl semimetal at $z>0$ with $\A_5\parallel\hat\z$ as depicted in Fig.~\ref{fig:setup}, where an electromagnetic wave with $\k\parallel\hat\z$ normal to the surface is subjected to the boundary condition so as to vanish at $z\to+\infty$.
Because its wave number must have a positive imaginary part corresponding to the upper sign in Eq.~(\ref{eq:dispersion}), the vector potential at $z>0$ is provided by Eq.~(\ref{eq:potential}) with
\begin{align}\label{eq:solution}
k_\lambda(\omega) = \frac{\bar\sigma_M}{2} + \sqrt{\left(\frac{\bar\sigma_M}{2}\right)^2
+ i\bar\sigma_E\omega - \bar\sigma_H\omega + \frac{\omega^2}{v^2}}.
\end{align}
On the other hand, the vector potential at $z<0$ is a superposition of incident and reflected waves as
\begin{align}
\A(t,\r)|_{z<0}
&= A_+(\hat\x + \lambda i\hat\y)\,e^{-i\omega t+i(\omega/c)z} \notag\\
&\quad + A_-(\hat\x + \lambda i\hat\y)\,e^{-i\omega t-i(\omega/c)z}.
\end{align}
The boundary conditions at $z=0$,
\begin{align}
\E(t,\r)|_{z\to-0} &= \E(t,\r)|_{z\to+0}, \\
\frac{\B(t,\r)|_{z\to-0}}{\mu_0} &= \frac{\B(t,\r)|_{z\to+0}}{\mu},
\end{align}
then lead to
\begin{align}
A_\pm = \frac{\mu_r\omega\pm c\,k}{2\mu_r\omega}
A_{\omega,k}^\lambda\bigg|_{k=k_\lambda(\omega)}
\end{align}
with $\mu_r=\mu/\mu_0$ being the relative permeability.
Therefore, the reflectance for given frequency $\omega$ and helicity $\lambda$ is found to be
\begin{align}\label{eq:reflectance}
R_\lambda(\omega) = \left|\frac{A_-}{A_+}\right|^2
= \left|\frac{\mu_r\omega-c\,k_\lambda(\omega)}{\mu_r\omega+c\,k_\lambda(\omega)}\right|^2.
\end{align}

By definition, $k_\lambda(\omega)$ in Eq.~(\ref{eq:solution}) has the positive imaginary part according to the boundary condition at $z\to+\infty$.
When it has a positive real part, the electromagnetic wave induced at $z>0$ is the usual one damping exponentially toward the direction of its propagation ($+\hat\z$) and thus leads to $R_\lambda(\omega)<1$.
This is the case for $\bar\sigma_M>0$ as well as for $\bar\sigma_M<0$ with $\bar\sigma_H<0$.
However, for $\bar\sigma_M<0$ with $\bar\sigma_H>0$, there exists the finite range of $0<\omega<\omega_0$ where $k_\lambda(\omega)$ has the negative real part as discussed above.
When this is the case, the electromagnetic wave triggered at $z>0$ is the anomalous one growing exponentially toward the direction of its propagation ($-\hat\z$) and thus leads to $R_\lambda(\omega)>1$.
Such anomalous reflection amplifying the circularly polarized light with selected helicity is nothing other than the manifestation of the chiral magnetic instability generating an unstable electromagnetic wave propagating toward $\mu_5\A_5$.
In the latter case, the zero-frequency limit of Eq.~(\ref{eq:reflectance}) reduces to the simple form of
\begin{align}
\lim_{\omega\to0}R_\lambda(\omega)
= \frac{(\mu_r\bar\sigma_M - c\,\bar\sigma_H)^2 + (c\,\bar\sigma_E)^2}
{(\mu_r\bar\sigma_M + c\,\bar\sigma_H)^2 + (c\,\bar\sigma_E)^2},
\end{align}
which obviously exceeds unity for $\bar\sigma_M\bar\sigma_H<0$ and is maximized at $\bar\sigma_M=-c\sqrt{\bar\sigma_E^2+\bar\sigma_H^2}/\mu_r$ for given $\bar\sigma_E$ and $\bar\sigma_H$~\cite{zero-frequency}.

\begin{figure}[t]
\includegraphics[width=0.9\columnwidth]{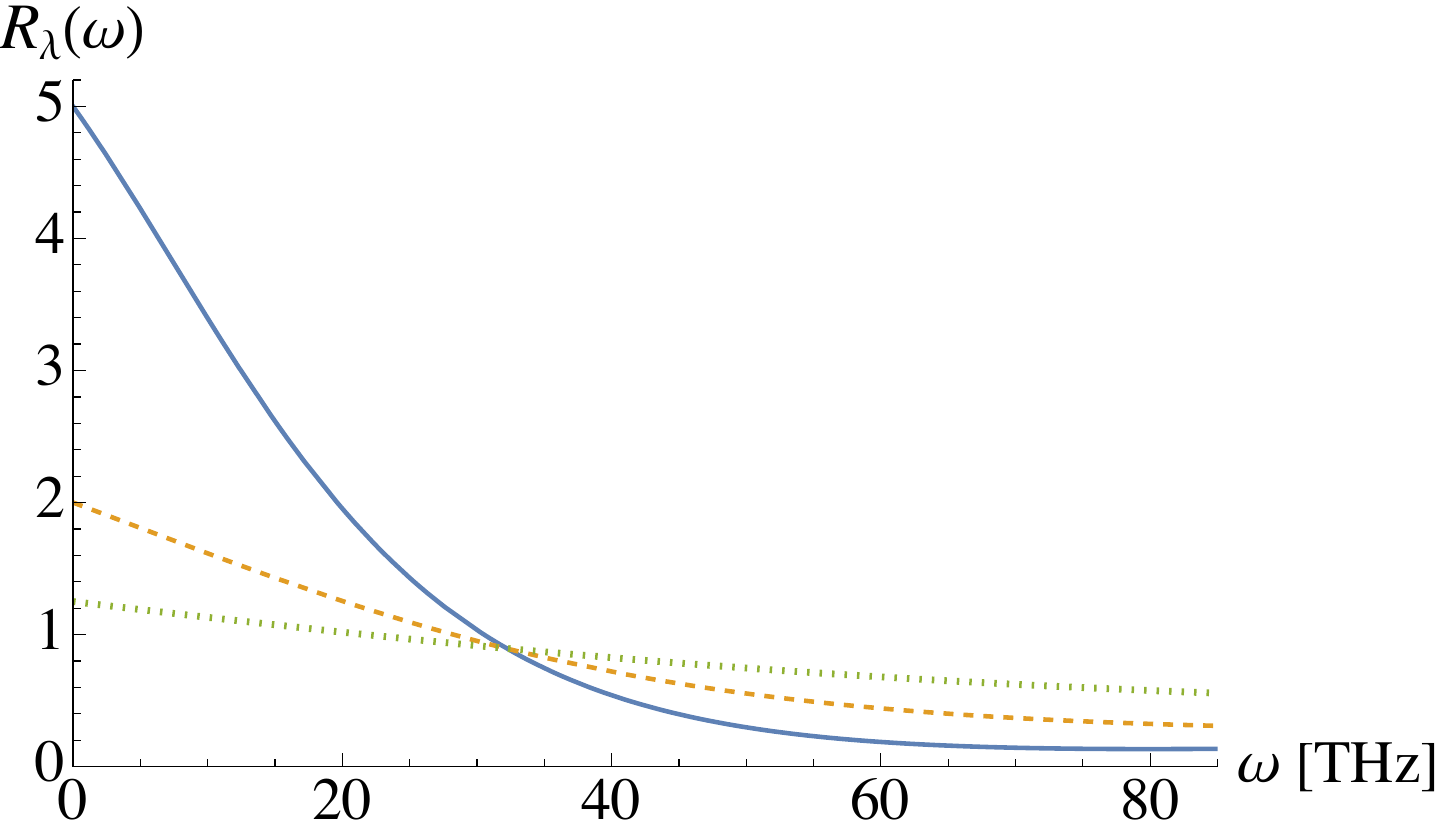}
\caption{\label{fig:reflectance}
Reflectances from Eq.~(\ref{eq:reflectance}) for $\lambda=-$ and $\hat\A_5=-\hat\z$ as functions of frequency with the parameter choice of $v=c/6$, $\mu_r=1$, $\sigma_H=10^4~\Omega^{-1}\,\m^{-1}$, and $\sigma_M=c\,\sigma_H$.
The solid, dashed, and dotted curves correspond to $\sigma_E=\sigma_H$, $2\sigma_H$, and $4\sigma_H$, respectively, for which the resulting reflectances exceed unity at $\omega<31$, 28, and 22~THz.}
\end{figure}

In order to quantify our prediction and discuss its experimental feasibility at the level of order estimate, we fix the anomalous Hall conductivity at $\sigma_H\approx10^4~\Omega^{-1}\,\m^{-1}$ motivated by the Weyl node separation of $|2e\A_5|/\hbar\approx10^9~\m^{-1}$~\cite{Xu:2015,Zhang:2016}.
Because the chiral magnetic conductivity can, in principle, be controlled by applied electric and magnetic fields, $\sigma_M=c\,\sigma_H/\mu_r$ is optimistically chosen so as to nearly maximize the reflectance at zero frequency for $\lambda=-$ and $\hat\A_5=-\hat\z$.
Then, $R_\lambda(0)=1+4\sigma_H^2/\sigma_E^2$ turns larger if the electric conductivity is smaller, and we employ threefold $\sigma_E=\sigma_H$, $2\sigma_H$, and $4\sigma_H$ corresponding to different temperatures, which lead to $R_\lambda(0)=5$, 2, and 1.25, respectively.
With the further choice of $v\approx c/6$ and $\mu_r\approx1$~\cite{Buckeridge:2016}, the resulting reflectances are presented in Fig.~\ref{fig:reflectance} as functions of frequency in units of terahertz.
Although the reflectance for the positive helicity (not presented) is always less than unity, that for the negative helicity is indeed found to exceed unity in the finite range of frequency below $\omega_0\sim10$~THz.
The roles of positive and negative helicities are exchanged when the signs of $\mu_5$ and $\A_5$ are simultaneously flipped.
We also note that our formula in Eq.~(\ref{eq:reflectance}) obtained with Eq.~(\ref{eq:current}) is valid only for $\omega\ll1/\tau\approx1$~THz~\cite{Zhang:2016}, which actually sets the upper limit of frequency still within the experimental reach.
Therefore, the significant amplification of the circularly polarized light is expected once $\sigma_E\lesssim\sigma_H\approx\mu_r\sigma_M/c$ is achieved.
However, the reality is $\mu_r\sigma_M/c\ll\sigma_H\ll\sigma_E$ as indicated by $\sigma_E\approx10^7~\Omega^{-1}\,\m^{-1}$ and $\sigma_M/c\approx1~\Omega^{-1}\,\m^{-1}$~\cite{Zhang:2016,Zhang:footnote}, for which the anomalous reflectance reading $R_\lambda(\omega)\approx1+10^{-10}$ below $\omega_0\approx10$~Hz is practically unobservable.

\section{Summary}
In summary, the chiral magnetic instability predicts unstable electromagnetic waves at low frequency and long wavelength as a combined consequence of the chiral magnetic effect and the anomalous Hall effect.
We showed that such chiral magnetic instability manifests itself as anomalous reflectance of the surface of Weyl semimetal in its nonequilibrium steady state with pumped axial charge.
The reflectance was found to exceed unity in a finite range of frequency for a circularly polarized light incident along the direction of Weyl node separation.
Therefore, the pumped Weyl semimetal is to serve as a ``chiral light amplifier'' with helicity selectivity and its performance can, in principle, be significant depending on electric, chiral magnetic, and anomalous Hall conductivities.
Although the observable performance is currently unrealistic, our finding hopefully stimulates further studies for the functional chiral light amplifier with pumped Weyl semimetals or beyond.

\acknowledgments
This work was supported by JSPS KAKENHI Grants No.\ JP18H05405 and No.\ JP21K03384.


\begin{thebibliography}{99}
\bibitem{zero-frequency}
The anomalous reflectance in the zero-frequency limit can be generalized to the case of $\hat\A_5\cdot\hat\z=\cos\theta$, where its formula with $\bar\sigma_H\equiv\lambda\mu\sigma_H$ redefined is provided by

$\displaystyle\lim_{\omega\to0}R_\lambda(\omega)
= \frac{(\mu_r\bar\sigma_M - c\,\bar\sigma_H\cos\theta)^2
+ c^2(\bar\sigma_E^2 + \bar\sigma_H^2\sin^2\theta)}
{(\mu_r\bar\sigma_M + c\,\bar\sigma_H\cos\theta)^2
+ c^2(\bar\sigma_E^2 + \bar\sigma_H^2\sin^2\theta)}.$\\
\end{thebibliography}
\end{document}